\documentstyle[12pt,aasms4]{article}
\def\lapprox{\hbox{\lower .8ex\hbox{$\,\buildrel < \over\sim\,$}}}
\def\gapprox{\hbox{\lower .8ex\hbox{$\,\buildrel > \over\sim\,$}}}

\begin{document}

\title
{Inhomogeneous Big Bang Nucleosynthesis with Late--Decaying Massive
Particles}

\author
{J. L\'opez--Su\'arez\altaffilmark{1} and R. Canal\altaffilmark{1,2}}

\altaffiltext
{1}{Department of Astronomy, University of Barcelona, Mart\'\i\ i Franqu\'es
1, E--08028 Barcelona, Spain. E--mail: jlopez@mizar.am.ub.es,
ramon@mizar.am.ub.es}

\altaffiltext
{2}{Institut d'Estudis Espacials de Catalunya/UB, Edif. Nexus--104, Gran
Capit\`a 2--4, 08034 Barcelona, Spain}

\slugcomment{{\it Running title:} Big Bang Nucleosynthesis}

\begin{abstract}
We investigate the possibility of accounting for the currently inferred
primordial abundances of D, $^{3}$He, $^{4}$He, and $^{7}$Li by big bang
nucleosynthesis in the presence of baryon density inhomogeneities plus the
effects of late--decaying massive particles (X), and we explore the allowed
range of baryonic fraction of the closure density $\Omega_{b}$ in such
context. We find that, depending on the parameters of this composite model
(characteristic size and density contrast of the inhomogeneities;
mass--density, lifetime, and effective baryon number in the decay of the X
particles), values as high as $\Omega_{b}h_{50}^{2}\simeq 0.25-0.35$ could be
compatible with the primordial abundances of the light nuclides. We include
diffusion of neutrons and protons at all stages, and we consider the
contribution of the X particles to the energy density, the entropy production
by their decay, the possibility that the X--products could photodissociate
the light nuclei produced during the previous stages of nucleosynthesis, and
also the possibility that the decay products of the X--particles would
include a substantial fraction of hadrons. Specific predictions for the
primordial abundance of Be are made.
\end{abstract}


\section{Introduction}
Standard homogeneous big bang nucleosynthesis could have produced the
observationally inferred primordial abundances of D, $^{3}$He, $^{4}$He, and
$^{7}$Li, provided that the baryon fraction of the cosmic closure density
$\Omega_{b}$ would lie in the range:

$$0.04\lapprox \Omega_{b}h_{50}^{2}\lapprox 0.08\eqno(1)$$

\noindent
where $h_{50}$ is the Hubble constant in units of 50 km s$^{-1}$ Mpc$^{-1}$
(Walker et al. 1991; Copi, Schramm, \& Turner 1995). For the long--time most
favored cosmological model, a flat Universe with $\Omega_{M} = 1$ and
$\Omega_{\Lambda} = 0$ (those being, respectively, the fractional
contributions of matter and vacuum energy densities to the closure density),
the upper bound to $\Omega_{b}$ would mean that most matter in the Universe
should be in nonbaryonic form. Given the far--reaching implications of the
dominance of nonbaryonic dark matter, possible alternatives to homogeneous
big bang nucleosynthesis have been explored, especially during the last 15
years or so.

The suggestion that the quark--hadron phase transition might be first--order
and generate baryon inhomogeneities (Witten 1984) led to the calculation of
the possible effects on primordial nucleosynthesis (Applegate \& Hogan 1985;
Applegate, Hogan, \& Scherrer 1987; Malaney \& Fowler 1988). The goal was to
see whether inhomogeneous big bang nucleosynthesis with $\Omega_{b} = 1$
might account for the primordial light--element abundances. Besides a
first--order quark--hadron phase transition, other mechanisms might also
generate baryon inhomogeneities. Much of the work in this line is reviewed by
Malaney \& Mathews (1993). However, the recent studies, treating accurately
the coupling between  baryon diffusion and nucleosynthesis, show that the
upper limit on $\Omega_{b}$ set by the light--element abundances does not
significantly differ from that obtained for homogeneous big bang
nucleosynthesis (Mathews, Schramm, \& Meyer 1993; Thomas et al. 1994). This
last conclusion, though, has very recently been challenged by Orito et al.
(1997), who explore the dependence of primordial nucleosynthesis on the
geometry of baryon inhomogeneities and find that cylindrical geometry might
allow to satisfy the observational constraints with baryon fractions as high
as $\Omega_{b}h_{50}^{2}\lapprox 0.2$.

A different approach has been to explore the possible modifications of the
yields from homogeneous big bang nucleosynthesis by the effects of the decay
of unstable massive particles ($M\gapprox few\ GeV$), produced at earlier
stages in the evolution of the Universe and with half--lives longer than the
standard nucleosynthesis epoch ($\tau_{x}\gapprox  10^{4}\ s$) (Audouze,
Lindley, \& Silk 1985; Dom\'{\i}nguez--Tenreiro 1987; Yepes \&
Dom\'{\i}nguez--Tenreiro 1988; Dimopoulos et al. 1988). Gravitinos produced
during reheating at the end of inflation are a possible example of such
particles. In Dimopoulos et al. (1988), the emphasis is put on the resulting
hadron cascade. The main problem encountered in this model is the predicted
overproduction of $^{6}$Li: $^{6}Li/^{7}Li\gg 1$, whereas observations show
$^{6}Li/^{7}Li\lapprox 0.1$.

Although they have only been considered separately, baryon inhomogeneities
and the presence of unstable massive particles decaying when the Universe has
already cooled down below $T_{9}\simeq 0.4$ are by no means mutually
exclusive. Here we explore their combined effects on the primordial
abundances of the light elements. The parameter space now has, of course, a
dimension which is the sum of those for the two separate cases:
characteristic size and density contrast of the inhomogeneities,
mass--density, lifetime, and mode of decay of the massive particles. We find
that there are regions in such extended parameter space where values of
$\Omega_{b}$ as high as $\Omega_{b}h_{50}^{2}\simeq 0.35$ would still be
compatible with the primordial abundances of the light nuclides inferred from
observations. Such values of $\Omega_{b}$ are of the same order as the low
values for $\Omega_{M}$ currently derived from a variety of sources,
including high--redshift supernova searches (Perlmutter et al. 1998;
Garnavich et al. 1998). Our results thus suggest again the possibility that
all the matter in the Universe could be baryonic.

On the other hand, recent determinations of the D abundance in high--redshift
QSO absorbers, when confronted with the currently inferred primordial
$^{4}$He abundance, might be in conflict with the predictions of standard,
homogeneous big bang nucleosynthesis for $N_{\nu} = 3$ (Steigman 1998): the
``low'' high--redshift D abundances (which appear more reliable) would
indicate too high a value of $\Omega_{b}$ to be compatible with that
corresponding to the $^{4}$He abundance. Since it is hard to tell whether
this conflict points to new physics or just to systematic errors in the
derivation of abundances, Steigman, Hata, \& Felten (1998) have discarded the
constraint on $\Omega_{b}$ from standard big bang nucleosynthesis and turned
to other observational constraints to determine the key cosmological
parameters. The results from our composite model, by showing how minor
deviations from the standard hypotheses can produce agreement with the
primordial abundances inferred from observations, support that attitude.
Besides, as we will see, the combined effects of inhomogeneities plus
late--decaying particles might solve the conflict between D and $^{4}$He
abundances.

\section{Model, Results, and Discussion}

In the present model, nucleosynthesis first takes place in a Universe with
baryon inhomogeneities. Later, when the temperatures are low enough for the
chemical abundances to be frozen, unstable massive particles start to decay
producing both hadronic and electromagnetic showers which alter the
abundances of the light elements resulting from the previous stage.

For the inhomogeneities, we consider a simple model consisting of two types
of zones, one with high density and the other with low density, characterized
by their respective volume fractions $f_{v}$ and $1 - f_{v}$, and by the
density contrast $R$ between the two zones (see Thomas et al. 1993). The
inhomogeneities, produced at some earlier stage, lead to differential
diffusion of protons and neutrons  when the temperature, in the expanding
Universe, drops below $T\simeq 1\ MeV$ and protons and neutrons are no longer
in equilibrium (Applegate \& Hogan 1985). Then, due to the much longer mean
free path of the neutrons, the initial baryon inhomogeneities transform into
variations of the local neutron/proton ratio (Applegate, Hogan, \& Scherrer
1987, 1988; Rauscher et al. 1994). When nucleosynthesis starts at
$T_{9}\simeq 1$, neutrons are already uniformly distributed whereas the
protons retain the spatial distribution they had at weak decoupling.
Nucleosynthesis thus takes place in two different types of zones: the
proton--rich and the neutron--rich ones. Since almost all neutrons end up
into $^{4}$He, with the rapid growth of the abundance of this nuclide protons
become exhausted in the neutron--rich zones and the same occurs with neutrons
in the proton--rich zones. A density contrast thus appears again and neutrons
start to diffuse back from the neutron--rich zones into the proton--rich
ones. We have coupled neutron diffusion with the nuclear reaction network as
in Rauscher et al. (1994).

On the other hand, we have the X--particles, with masses $m_{x}$ and
half--lives $\tau_{x}$. The particles are massive ($m_{x} > 10\ GeV$) and we
consider only half--lives in the interval $10^{4}\ s\leq \tau_{x}\leq 10^{7}\
s$. The former means that they are nonrelativistic well before the start of
nucleosynthesis. The lower limit to the half--life implies that the
thermonuclear reaction rates have dropped to zero before the X--particles
start to disintegrate, while the upper limit ensures that their decays leave
no imprint on the cosmic background radiation. Prior to decaying, the
X--particles just give a contribution $\rho_{x}$ to the matter--energy
density. We define $r$ as the number ratio of the X--particles to photons at
some fiducial temperature $T_{0}$. Here we follow the history of cosmic
expansion starting at $T_{0} = 10^{12}\ K$. The product:

$$rm_{x} \equiv \left({n_{x}m_{x}\over n_{\gamma}}\right)_{T_{0}}\eqno(2)$$

\noindent
is one of the parameters of the model. When a particle of mass higher than a
few GeV decays during the keV era, a fraction of its energy goes into
electromagnetic decay products while another fraction goes into hadrons. Both
kinds of products give rise to showers as they thermalize with the background
plasma. The baryonic decay products interact with the ambient protons and
$\alpha$--particles and initiate chains of nuclear reactions that lead to
production of the light nuclides D, $^{3}$He, $^{6}$Li, $^{7}$Li, and to
destruction of some $^{4}$He (Dimopoulos et al. 1988). The electromagnetic
decays consist of the injection of high--energy photons, electrons, and
positrons into the background plasma. The high--energy photons dissociate the
light nuclei. The average number of baryons $\nu_{B}$ produced in the decay
of the X--particles has been calculated by Schwitters (1983). The numbers
$\xi_{i}$ of nuclei $i$ and neutrons produced by the disintegration of each
X--particle can be calculated by modeling the hadronic cascades (Dimopoulos
et al. 1988, 1989).

In order to calculate the final abundance for each light nuclide $i$,
including the effects of inhomogeneities plus X--particle decay, we
integrate, starting at $T_{0} = 10^{12}\ K$, the following set of
differential equations:

$${dY_{i}\over dt} = \pm\delta_{in}\kappa Y_{n} +{\eta_{0} rm_{x}\over
\tau_{x}}exp(-t/\tau_{x})\left[\xi_{i}r^{*}_{B} - {Y_{i}\over Y_{p}}
\int_{Q_{i}}^{E_{max}} {\epsilon_{\gamma}(E) \sigma_{\gamma i}(E)\over
\sigma_{C}(E)}dE\right] + \left({dY_{i}\over dt}\right)_{std}\eqno(3)$$

\noindent
where the + sign corresponds to the proton--rich zone, the -- sign to the
neutron--rich one, $\kappa$ is the diffusion rate of neutrons, $\eta_{0}$ is
the initial $n_{b}/n_{\gamma}$ ratio, $r_{B}^{*}$ is the effective baryonic
branching ratio of the X--decays, $\epsilon_{\gamma}(E)$ is the photon
spectrum produced by disintegration of the X--particles, $\sigma_{\gamma
i}(E)$ is the photodissociation cross--section, and $\sigma_{C}(E)$ is the
Coulomb scattering cross--section. The integrals extend from the
photodissociation threshold energy $Q_{i}$ to $E_{max} = m_{e}^{2}/(25T)$
(which thus increases with time). Subscripts $n$ and $p$ refer to neutrons
and protons, respectively, and the subscript $std$ indicates the standard
contribution of thermonuclear reactions to $dY_{i}/dt$. In order to deal with
diffusion in the way indicated in (3), the nuclear abundances $Y_{i}\equiv
N_{i}/N_{b}$ are taken relative to the total number of baryons in the given
volume before neutron diffusion sets in, as in Rauscher et al. (1994). The
neutron diffusion rate is given by:

$$\kappa = {4.2\times 10^{4}\over (d/a)} T_{9}^{5/4}(1 + 0.716T_{9})^{1/2}\
s^{-1}\eqno(4)$$

\noindent
where $d/a$ (in $cm\ MeV$) is the comoving length scale of the
inhomogeneities (Applegate, Hogan, \& Scherrer 1988). The effective baryon
branching ratio $r_{B}^{*}$ in (3) is another parameter of the model. It
takes into account the dependence of the number of baryons produced by
disintegration of the X--particles on their mass $m_{X}$ together with the
dependence of the yields $\xi_{i}$ on the kinetic energies of the primary
shower baryons:

$$r_{B}^{*} = \left({\nu_{B}\over 5}\right) r_{B} F\eqno(5)$$

\noindent
where $r_{B}$ is the true baryonic branching ratio, and $F$ incorporates the
dependence of the yields $\xi_{i}$ on the kinetic energy of the primary
shower baryons (our ``standard'' $\xi_{i}$ have been calculated for $m_{x} =
1\ TeV$, $\nu_{B} = 5$, and $E_{kin} = 5\ GeV$). The photons produced by the
decay of the X--particles have a spectrum:

$$\epsilon_{\gamma}(E) = \left({m_{x}\over
2E_{max}^{1/2}}\right)E^{-3/2}\eqno(6)$$

Our model, therefore, has five extra free parameters in addition to
$\eta_{0}$: the volume fraction $f_{v}$ and the density contrast $R$
characterizing the inhomogeneities (but we also vary the comoving length
scale $d/a$ governing neutron diffusion), the mass $m_{x}$ (in fact the
product $rm_{x}$) and the half--life $\tau_{x}$ of the X--particles, and the
effective baryon ratio $r_{B}^{*}$ in the X--particle decays. We thus solve
the equations (5) for different combinations of those parameters, searching
for the regions in the parameter space where the predicted primordial
abundances of the light nuclides are compatible with those inferred from
observations. The nuclear reaction network consists of 161 thermonuclear
reactions, plus 20 reactions involved in the hadron cascades, plus 15
photodisintegration reactions. In following the expansion of the Universe we
take into account the contribution of the X--particles to the total energy
density and the entropy change due to their disintegration (assuming that the
X--decay products thermalize on a time scale much shorter than the expansion
time scale). The cross--section sources and a more detailed description of
the model will be given elsewhere (see also L\'opez--Su\'arez 1997).

The set of primordial abundances to be fitted is the following:

$$1.1\times 10^{-5}\leq\left({D\over H}\right)_{p}\leq
2.5\times10^{-4}\eqno(7a)$$

$$3.3\times 10^{-5}\leq\left({D + ^{3}He\over H}\right)_{p}\leq
2.5\times10^{-4}\eqno(7b)$$

$$0.21\leq X(^{4}He)_{p}\leq 0.24\eqno(7c)$$

$$1.1\times10^{-10}\leq \left({^{7}Li\over H}\right)_{p}\leq
2.6\times10^{-9}\eqno(7d)$$

$$\left({^{6}Li\over ^{7}Li}\right)_{p} < 1\eqno(7e)$$

The limits in (7a--e) are adopted, respectively, from McCullough (1992),
Geiss \& Reeves (1972), Pagel \& Kazlaukas (1992), Krauss \& Kernan (1994),
and Smith, Lambert, \& Nissen (1996). In (7e) we have adopted a conservative
upper limit, taking account of the possibility that any primordial $^{6}$Li
might have been partially destroyed ($^{7}$Li remaining almost intact) in
metal--poor halo stars.

We have varied the volume fraction within the interval $0.01\leq f_{v}\leq
0.28$, and the density contrast within $50\leq R\leq 5000$. For the comoving
length scale $d/a$, we have tried the cases without diffusion and with
$10^{5.5}\ cm\ MeV\leq d/a\leq 10^{7.5}\ cm\ MeV$. For the parameters related
to the X--particles, $10^{-5}\ GeV\leq rm_{x}\leq 10^{3}\ GeV$,
$1.5\times10^{-12}\leq r_{B}^{*}\leq 1.5\times10^{-9}$, and $10^{4}\ s\leq
\tau_{x}\leq 10^{7}\ s$.

As an example of the results, in Figure 1 we show the dependence of the final
D abundance on $\tau_{x}$, for three different values of $f_{v}$ and fixed
values of $R$, $d/a$, $r^{*}_{B}$, $rm_{x}$, and $\eta_{f}$ (the final value
of the baryon to photon ratio). We see that for $5\times10^{5}\ s\lapprox
\tau_{x}\lapprox 6\times10^{5}\ s$ the resulting abundances fall within the
range allowed by observations. In Figure 2 we show the dependence on
$\tau_{x}$ of the final abundances of D, $^{3}$He, $^{4}$He, $^{6}$Li,
$^{7}$Li, and $^{9}$Be, for fixed value of $f_{v}$ and the same values of the
other parameters as in Figure 1. The two vertical dashed lines mark the
interval of values of $\tau_{x}$ compatible with all the observationally
inferred abundances of the light nuclides.

In summary, we obtain results which are compatible with the observations for
little diffusion ($d/a = 10^{7.5}\ cm\ MeV$), small abundances of the
X--particles ($rm_{x}\sim 10^{-5}\ GeV$), and modest numbers and energies of
the shower baryons ($1.5\times10^{-12}\leq r_{B}^{*}\leq 1.5\times
10^{-11}$). The density contrast between the two model zones must be $500\leq
R\leq 5000$, and the volume fraction $0.144\leq f_{v}\leq 0.192$. The
half--lives of the X--particles are $6.19\times10^{5}\ s\leq \tau_{x}\leq
7.43\times10^{5}\ s$, as illustrated in Figures 1 and 2. In that region of
the parameter space $18\leq \eta_{10}\leq 22$ ($\eta_{10}$ being here, as
ususal, $\eta$ in units of 10$^{-10}$). That corresponds to a baryon
fraction:

$$0.25\leq \Omega_{b}h_{50}^{2}\leq 0.35\eqno(8)$$

\noindent
in strong contrast with (1). A testable prediction of the present model is
the production of an appreciable amount of $^{9}$Be:

$$\left({^{9}Be\over H}\right)_{p}\sim 10^{-13}\eqno(9)$$

The production of $^{9}$Be is characteristic of inhomogeneous big bang models
(Malaney \& Fowler 1989; Jedamzik et al. 1994; Orito et al. 1997). The
current observational upper limit to the Be abundance (Duncan et al. 1997;
Garc\'{\i}a L\'opez et al. 1998) is of the same order as the values found
here. There is, however, the problem that $B/Be \sim 10$ almost down to
$[Fe/H] = -3$, and since our model predicts $B/Be < 1$, agreement would
require a drop in the ratio taking place below some still smaller
metallicity.

\section{Conclusions}

We have shown, by means of a simple model, that the combined effects on big
bang nucleosynthesis of baryon inhomogeneities plus the decay of unstable,
relatively long--lived massive particles, giving rise to both electromagnetic
and hadron cascades, might be to allow agreement with the primordial
light--element abundances inferred from observations for values of
$\Omega_{b}$ much higher that those allowed by standard, homogeneous
nucleosynthesis. The upper limit might be as high as
$\Omega_{b}h_{50}^{2}\simeq 0.35$. The values obtained here are of the same
order as the low $\Omega_{M}$ values now being derived from a variety of
sources and, therefore, they pose in new terms the question of whether all
matter in the Universe could be baryonic. A testable prediction of the model
is the production of a $^{9}$Be abundance that is of the order of current
observational upper limits.

On the other hand, in the parameter region of our model where there is
agreement between predicted and observationally inferred primordial
light--element abundances, given values of $\Omega_{b}$ (or, equivalently,
$\eta_{10}$) always predict ``low'' D abundances (in the sense of the
high--redshift abundances referred to in the Introduction), thus potentially
eliminating the conflict with the $^{4}$He abundance pointed out by Steigman
(1998).

The model presented here deals with inhomogeneities in a very simplified way.
A futher step will be to examine the effects of the geometry of the density
fluctuations on the outcome. Orito et al. (1997) have already shown that
cylindrical shell geometry alone (without the extra effects of late--decaying
particles) might allow $\Omega_{b}\lapprox 0.2$ (but for density contrasts
$R\sim 10^{6}$, much higher than those considered here). Another extension of
the model will be to consider particles with shorter half--lives, decaying at
the time when thermonuclear reactions are still taking place.

\clearpage

\begin{figure}[hbtp]
\centerline{\epsfysize15cm\epsfbox{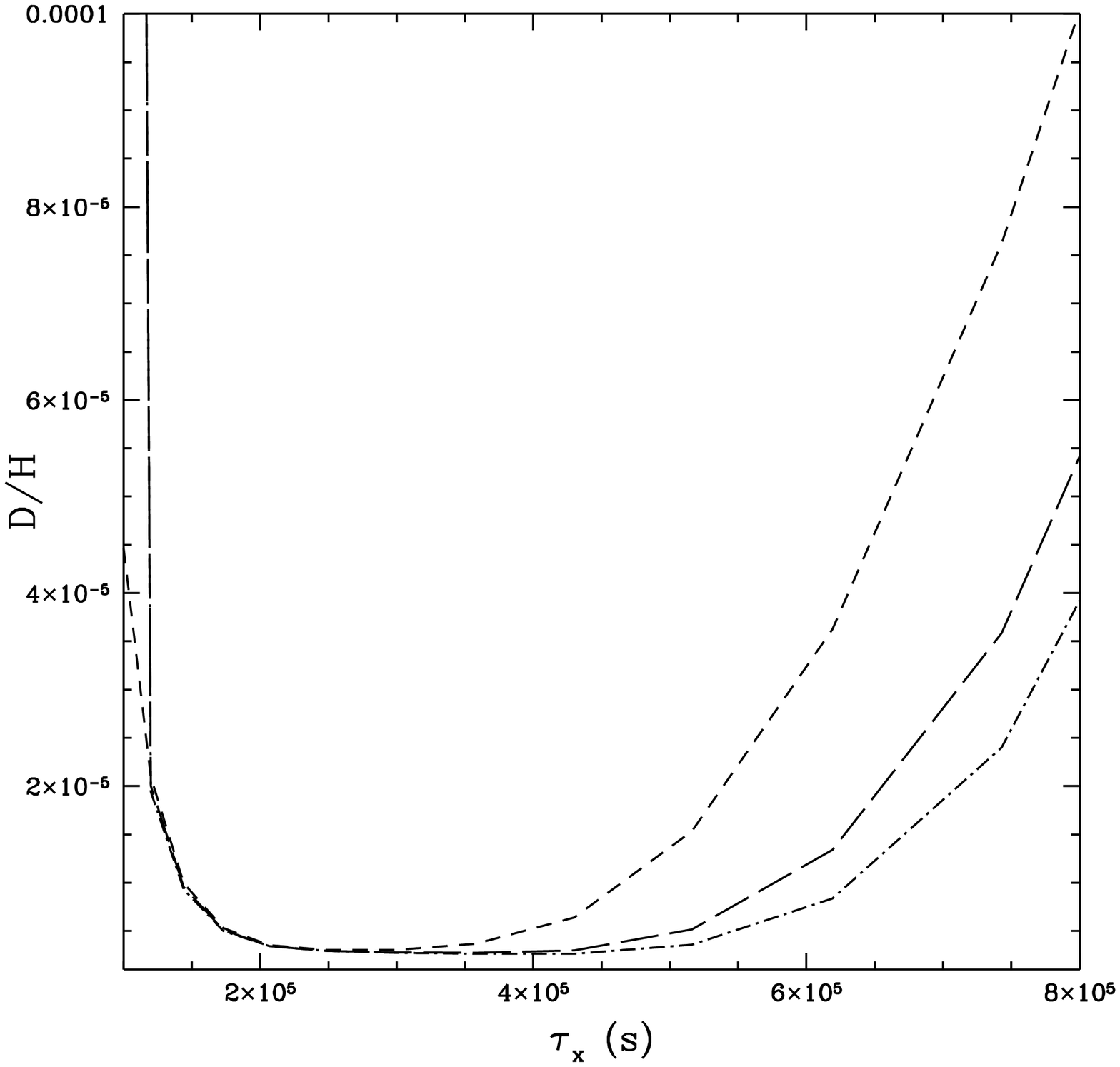}}
\nopagebreak[4]
\figcaption{Primordial D abundance as a function of the $\tau_{x}$, the
half--life of the X--particles, for three different values of the volume
fraction $f_{v}$: 0.074 (short--dashed line), 0.119 (long--dashed line), and
0.144 (dot--dashed line), and fixed values of the other parameters (see text
for the meaning of the different symbols).
\label{fig1}}
\end{figure}

\begin{figure}[hbtp]
\centerline{\epsfysize15cm\epsfbox{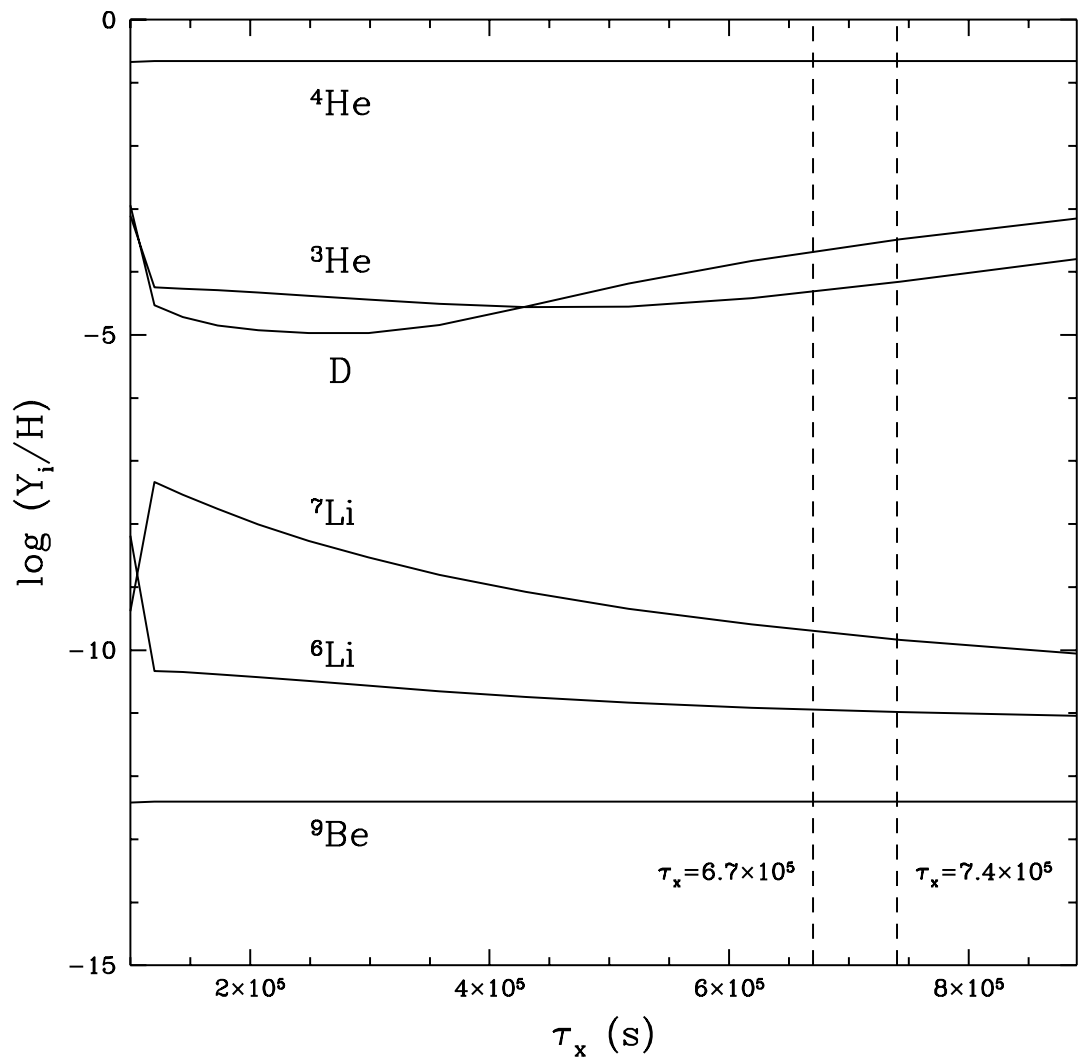}}
\nopagebreak[4]
\figcaption{Primordial abundances of the light nuclides D, $^{3}$He,
$^{4}$He, $^{6}$Li, $^{7}$Li, and $^{9}$Be, as a function of the half--life
of the X--particles, $\tau_{x}$, for fixed values of the other parameters:
$\eta_{10, f} = 18$, $f_{v} = 0.144$, $R = 5000$, $d/a = 10^{7.5}$, $rm_{x} =
10^{-5}$, and $r_{B}^{*} = 1.5\times10^{-12}$ (see text for the meaning of
the different symbols). The two vertical dashed lines mark the boundaries of
the interval where the predicted abundances are compatible with those
observationally inferred.
\label{fig2}}
\end{figure}

\end{document}